# COMPUTING CLASSICAL ORBITAL ELEMENTS WITH IMPROVED EFFICIENCY AND ACCURACY

R. Flores[a] and E. Fantino[b]*

[a] Department of Aerospace Engineering, Khalifa University of Science and Technology, P.O. Box 127788, Abu Dhabi (United Arab Emirates). Email: robertomaurice.floresleroux@ku.ac.ae

[b] Department of Aerospace Engineering, Khalifa University of Science and Technology, P.O. Box 127788, Abu Dhabi (United Arab Emirates). Email: elena.fantino@ku.ac.ae

* Corresponding author: elena.fantino@ku.ac.ae

Abstract

*This paper reviews the standard algorithm for converting spacecraft state vectors to Keplerian orbital elements with a focus on its computer implementation. It analyzes the shortcomings of the scheme as described in the literature, and proposes changes to the implementation to address orbits of arbitrary eccentricity and inclination in a simple and robust way. It presents two coding strategies that simplify the program structure while improving the accuracy and speed of the transformation on modern computer architectures. Comprehensive numerical benchmarks demonstrate accuracy improvements by two orders of magnitude, together with a 40% reduction of computational cost relative to the standard implementation.*

## 1. Introduction

Transforming between spacecraft State Vector (SV) $\mathbf{x} = \{\mathbf{r},\mathbf{v}\}$ (position and velocity) and classical orbital elements (COE) $\mathbf{o} = \{a,e,i,\Omega,\omega,\theta\}$ (semimajor axis, eccentricity, inclination, longitude of the ascending node, argument of pericenter and true anomaly) (Bate, et al., 1971) is a recurring task in many orbital mechanics calculations. There are many high-precision propagators that operate with the state vector, but it is often the case that the orbital elements provide a more convenient description of the trajectory. This is due to their small changes over one revolution[1], which makes them attractive for applications like station-keeping (Fantino, et al., 2017), long-term orbital stability (Proietti, et al., 2021), and orbit transfer optimization (Fantino, et al., 2023). Moreover, the orbital elements provide an intuitive understanding of the orbit orientation and shape.

The algorithm to transform from SV to COE is considered an essential tool of celestial mechanics and, as such, it is ubiquitous in the reference literature (Bate, et al., 1971) (Prussing & Conway, 2013) (Tewari, 2007) (Vallado & McClain, 2007) (Curtis, 2020). Unlike the switch from COE to SV, which is straightforward and involves a fixed sequence of steps, the inverse transform must take into account the singularities in the definition of the orbital elements. For equatorial orbits, $\Omega$ is ill-defined. In circular orbits, $\omega$ also becomes degenerate. These singularities are especially relevant for applications where the equations of motion are formulated directly in terms of the orbital elements (Lara, et al., 2014). Different sets of elements have been proposed to work around this limitation. For a review of alternatives see (Hintz, 2008). The family of equinoctial elements was a first attempt to address the issues. It traces its origins to Lagrange's secular theory of planetary motion (Lagrange, 1781). The term "equinoctial elements" was coined by Arsenault et al. (Arsenault, et al., 1970) in 1970, and multiple variations of them can be found in the literature (Moulton, 1970) (Broucke & Cefola, 1972) (Chobotov, 2002) (Battin, 1999) (Giacaglia, 1977) (Nacozy & Dallas, 1977) (Cohen & Hubbard, 1962). Equinoctial elements still suffer from limitations, they are singular for parabolic orbits and cannot treat simultaneously prograde and retrograde equatorial orbits. A further refinement are the modified equinoctial elements (Walker, 1986) (Walker, et al., 1985). This work considers only the transformation from SV to COE, and focuses on improving its accuracy and efficiency in modern computers.

---

[1] Of course, the true anomaly experiences large variations, but it can be replaced easily with the time of periapsis or the true/mean anomaly at epoch, which vary slowly.

Software implementations commonly handle the singularities by means of separate branches of code for each case. This approach increases program size and the likelihood of coding errors, and, crucially, can be detrimental for computational performance in modern CPUs. Moreover, identifying which code path to take involves setting tolerances for eccentricity and inclination. This may compromise accuracy if the thresholds are not chosen carefully.

Current processor architectures are pipelined (Shen & Lipasti, 2005). Each instruction goes through several execution stages (one per clock cycle) before completion, while subsequent instructions are moving through the pipeline concurrently. It is often the case that the result of a previous instruction is required to decide which path of a branch must be taken. Waiting for the result to be ready (i.e., for the value of the condition to come out of the pipeline) would waste processor cycles (this is known as a "bubble"). Bubbles are mitigated with speculative execution assisted by a branch predictor (Smith, 1981). The hardware makes a guess of the likely path to follow and starts execution before the condition can be evaluated. In case the prediction turns out erroneous, the pipeline must be flushed, wasting the processing already done for the instructions in flight (Eyerman, et al., 2006). An effective strategy against performance degradation due to branch mispredictions is, unsurprisingly, branchless programming (Pikus, 2021).

This document presents two coding strategies that simplify the calculation of classical orbital elements and improve the accuracy of the results. Section 2 reviews a standard implementation, as found in the astrodynamics literature. It outlines the main shortcomings and strategies to mitigate them. Section 3 presents a branchless version that eliminates the need to contemplate coordinate system singularities. It enables robust, strictly linear (i.e., free from branching) execution, for improved precision and speed. Section 4 compares the accuracies of the original and branchless algorithms in the context of low-inclination and near-circular orbits. Section 5 introduces an alternative implementation for computer systems where the branchless approach is not suitable. It retains substantial improvements in accuracy, and reduces the number of branches (without completely eliminating them). Section 6 presents additional benchmarks. Finally, the main conclusions are drawn in section 7.

## 2. Standard programming approach

The most ubiquitous algorithm in the literature (which shall be denoted as AL1) follows the steps outlined in Fig. 1 to compute the orbital elements from the state vector $\mathbf{x}$. All vectors are expressed in an inertial Cartesian frame $\{x, y, z\}$ with axes along the $\{\hat{\mathbf{i}}, \hat{\mathbf{j}}, \hat{\mathbf{k}}\}$ directions. Hats denote unit vectors, for example: $\hat{\mathbf{r}} = \mathbf{r}/r$, where $r = \sqrt{\mathbf{r} \cdot \mathbf{r}}$. The reference plane for measuring inclination is xy and $\mu$ denotes the gravitational parameter. What follows is a very succinct description of the scheme, just to highlight potential pitfalls. For more details on the theoretical foundation, the reader is referred to existing literature, e.g. (Bate, et al., 1971). Henceforth, "Step i" shall denote the i-th block of the corresponding flowchart.

Step 1 of the algorithm computes the orbital angular momentum $\mathbf{h}$. The angle between $\mathbf{h}$ and the z axis yields the orbital inclination $i$. The direction of the ascending node $\hat{\mathbf{n}}$ is determined from the cross product $\hat{\mathbf{k}} \times \mathbf{h}$ (Step 4). However, because this product vanishes for equatorial orbits, a branch and an alternative code path (Steps 3-4) must be included to avoid division by zero when normalizing $\hat{\mathbf{n}}$. In the alternative path, the ascending node direction is set equal to the x axis. This choice is arbitrary, but it is acceptable because $\Omega$ is not well-defined for equatorial orbits. The scheme, as presented in the flowchart, uses the sign transfer function

$$\text{sign}(a, b) = \begin{cases} |a| & \text{if } b \geq 0 \\ -|a| & \text{otherwise} \end{cases}, \tag{2.1}$$

to determine the correct sign of $\Omega$. This function is available in some programming languages (e.g., Fortran). In case it is not supported, it can be replaced with a simple branch (with a potential performance penalty, however).

Step 5 computes the semimajor axis from the total energy balance, as well as the eccentricity vector $\mathbf{e}$. The latter points in the direction of the pericenter and its magnitude gives the orbital eccentricity (i.e., $e = \sqrt{\mathbf{e} \cdot \mathbf{e}}$). The angle $\widehat{\hat{\mathbf{n}}, \mathbf{e}}$ gives the argument of the pericenter and $\widehat{\mathbf{e}, \mathbf{r}}$ is the true anomaly (Step 8). The sign of $\omega$ is determined checking if the pericenter is above or below the reference plane. The sign of $\theta$ is chosen from the sign of the radial velocity. Because the eccentricity vector vanishes for circular orbits, a branch (Step 6) is required to assign the arbitrary value $\omega = 0$ and measure $\theta$ from the ascending node in such cases (Step 7).

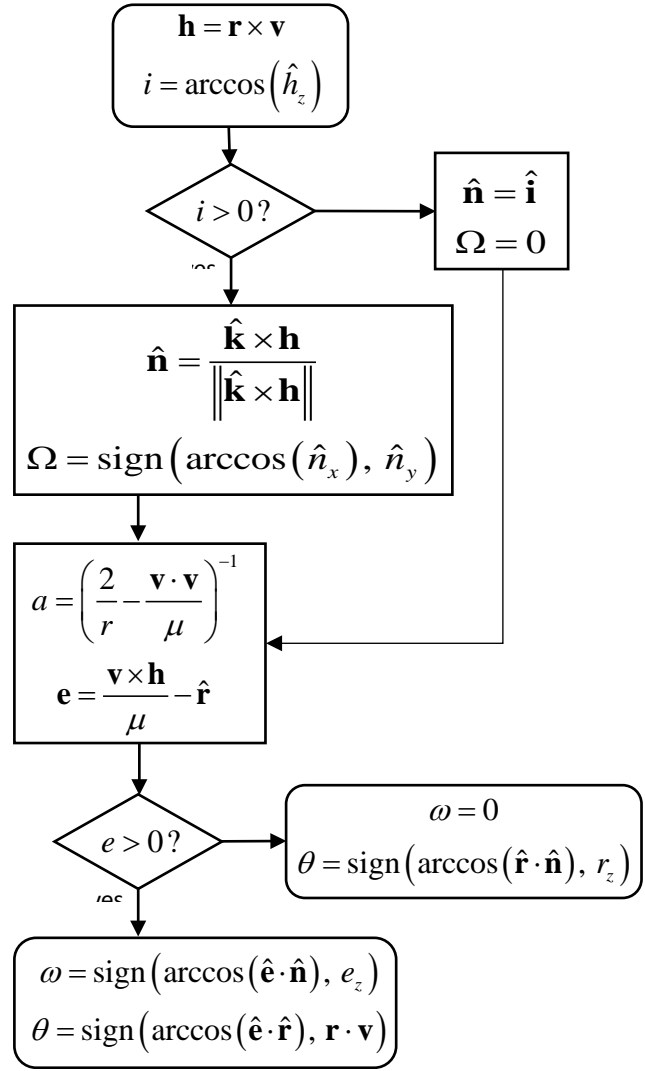

*Fig. 1 – Flowchart of traditional scheme (AL1).*

It is important to realize that, as described in the literature, AL1 does not even work for eccentric inclined orbits in practice, because Steps 7 and 8 hide an insidious pitfall. In exact arithmetic, one expects $\left|\hat{\mathbf{a}}\cdot\hat{\mathbf{b}}\right|\leq 1$ for any pair of unit vectors $\{\hat{\mathbf{a}},\hat{\mathbf{b}}\}$. However, finite precision arithmetic is subject to rounding errors. Thus, the magnitude of the dot product could exceed 1 if both vectors are quasi-parallel. This can happen, for example, in Step 8 when the spacecraft is very close to the apocenter or pericenter. The computer will trigger an exception if it tries to evaluate the arccosine of a value larger than 1, disrupting the calculation. While this occurrence may seem unlikely, if the elements are evaluated a large number of times, the problem will eventually arise[2]. To make matters worse, the circumstances where this issue is triggered depend on the exact sequence of operations used to compute the argument of the arccosine (this is typical with rounding-related problems). For example, these two apparently equivalent operations can yield completely different results (meaning one fails and the other does not):

---

[2] In fact, it was not possible to complete the tests in section 4 without addressing this problem first.

$$\arccos\left(\hat{\mathbf{a}} \cdot \hat{\mathbf{b}}\right) \;;\; \arccos\left(\frac{\mathbf{a} \cdot \mathbf{b}}{ab}\right). \tag{2.2}$$

A simple workaround is to replace the calculation of $\theta$ in Step 8 with:

$$\xi_1 = \max(\min(\hat{\mathbf{e}} \cdot \hat{\mathbf{r}}, 1), -1) \;;\; \theta = \text{sign}\left(\arccos\left(\xi_1\right), \mathbf{r} \cdot \mathbf{v}\right), \tag{2.3}$$

which, unfortunately, introduces additional latencies because $\theta$ cannot be evaluated until $\xi_1$ is available. The same applies to $\omega$ in Steps 7 and 8. On the other hand, computing the arccosine in Steps 1 and 4 is safe, because the argument is a single component of a unit vector. This operation is extremely robust numerically, and no special precautions are required.

The branches in Steps 2 and 6 are problematic because, aside from the potential performance penalty, conditions such as $i > 0$ and $e > 0$ do not fare well in the domain of finite precision arithmetic[3] (which always applies for numerical orbit propagation). Once the magnitude of the vectors becomes sufficiently small, the rounding errors inherent to floating point arithmetic can start dominating the calculations. While it may still be possible to normalize the vectors without triggering an overflow, the resulting direction can be erroneous due to this kind of contamination. A simple workaround is to set finite thresholds below which the alternative code paths are taken. However, these tolerances must be adjusted carefully. Too low a value does not prevent the rounding issues (and can even trigger an exception), while an overly large threshold will incorrectly characterize normal orbits as degenerate cases. Therefore, for the sake of performance and accuracy, it is highly desirable to write an algorithm that can run in a strictly linear way (i.e., branchless) and without tolerances that require tuning.

There are other subtle errors in AL1 that need to be fixed before it can be used for arbitrary orbits. Note that if the inclination vanishes, the sign of $\omega$ in Step 8 cannot be determined from the vertical component of the eccentricity vector. Instead, it is given by the sign of the *y* component of $\mathbf{e}$ (because in this case the reference direction for $\omega$ is the *x* axis). Similarly, in Step 7, if the orbit is equatorial, the sign of the true anomaly must be given by the *y* component of $\mathbf{r}$. Furthermore, one must consider if the orbit is prograde or retrograde when choosing the signs. Also, to handle retrograde orbits correctly, the actual condition that must be tested in Step 2 is $i > 0$ AND $i < \pi$ (it shall be written as $i \in {]0,\, \pi[}$ to save space in the flowcharts). The improved classical algorithm, including an eccentricity threshold $\left(e_{thr}\right)$ in Step 6 to detect quasi-circular obits, is summarized in Fig. 2.

---

[3] Due to a lucky coincidence, explained in section 3, the condition $i > 0$ is acceptable in this case.

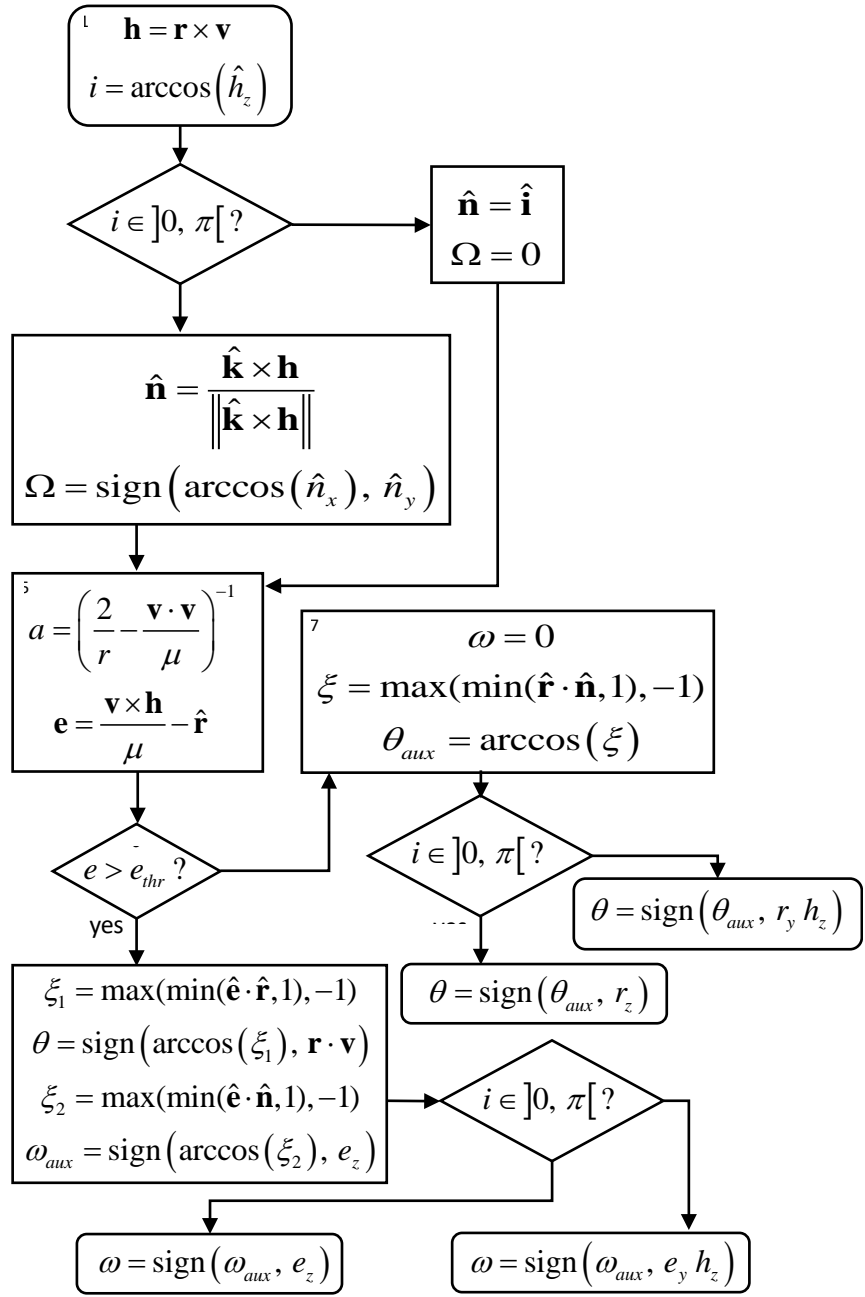

*Fig. 2 - Fully-functional version of traditional scheme (AL2).*

While the scheme in Fig. 2 (AL2 henceforth) addresses many issues from the original formulation, it is clearly more convoluted. This increases the likelihood of coding errors, and can negatively impact performance. It also requires setting an appropriate value for $e_{thr}$, which shall be discussed in section 4.1.

To close this section, the reader should note that Step 5 of AL1 and AL2 can trigger an exception —overflow, infinity or not a number (NaN), depending on the specifics of the computer, compiler and data— for near-parabolic orbits. A parabolic orbit, by definition, has infinite semimajor axis, so this is more a limitation of the COE themselves than a software issue. For most applications this edge case does not require special attention. For example, in double precision floating-point arithmetic it happens if $a \geq 1.8 \cdot 10^{308}$ (IEE, 2019). Note that, because the semimajor axis is not used in subsequent steps, the exception will not affect the computation of the other elements, and can be detected and treated a posteriori. For applications dealing with quasi-parabolic orbits, a simple workaround is to compute $a^{-1}$ instead, which is finite. This edge case falls outside the scope of the work, and will not receive further consideration. The classical scheme is directly applicable to hyperbolic orbits $(e>1)$, where it returns a negative semimajor axis, as per standard conventions.

## 3. A branchless general code path

### 3.1 Background

It is possible to improve the accuracy of the classical scheme, while reducing the branch penalties, by using the ATAN2 function ("partial arctangent" in computer parlance). It is widely available in modern compilers (e.g., invoking *math.h* in the C language). A call to ATAN2(*y*,*x*) returns the argument of the complex number $x+iy$ in the $[-\pi,\pi]$ range. Crucially, when the ATAN2 implementation conforms to the IEEE-754 floating-point standard (IEE, 2019), it deals cleanly with vanishing arguments

Furthermore, ATAN2 is a native instruction in all PC-compatible CPUs from Intel and AMD (x86 architecture). They power most laptops, desktops and workstations, as well as a majority of scientific HPC clusters. ATAN2's opcode (assembly instruction) is FPATAN (Intel Corp., 2024). Having direct hardware support can improve performance relative to software emulation.

Interestingly, mentions of the partial arctangent for transforming SV to COE can be found in the literature from the early 90s (Danby, 1992). However, the driving factor is completely different. The only inverse trigonometric function supported by the author's language of choice was ATAN. He was effectively forced to work with arctangents. He mentions ATAN2 (in systems where it is supported, which, unfortunately, was not his case) as a convenient way of computing $\arctan(y/x)$ and returning a value in the $[-\pi,\pi]$ range, instead of using an ATAN call (which operates inside $[-\pi/2,\pi/2]$), and then adjusting the quadrant depending on the signs of $x$ and $y$. The author explicitly states that the basic algorithm he presents is not suitable for circular or equatorial orbits, and that additional code paths (i.e., branches) are required to treat these cases. This algorithm (Danby, 1992) illustrate an alternative approach to the problem, and provides inspiration for developing the branchless scheme. Its steps (without provisions for circular or equatorial orbits) are[4]:

1. Compute $\mathbf{h}=\mathbf{r}\times\mathbf{v}$ and $\hat{\mathbf{n}}=\dfrac{\hat{\mathbf{k}}\times\mathbf{h}}{\left\|\hat{\mathbf{k}}\times\mathbf{h}\right\|}$

2. $i=\mathrm{ATAN2}\left(\sqrt{h_x^2+h_y^2}\ ,\ h_z\right)$

---

[4] The last step has been slightly modified from the original, where the time of passage through pericenter was used instead of the true anomaly.

3. $\Omega = \text{ATAN2}(\hat{n}_y, \hat{n}_x)$

4. Compute $a = \left(\dfrac{2}{r} - \dfrac{\mathbf{v} \cdot \mathbf{v}}{\mu}\right)^{-1}$ and $\mathbf{e} = \dfrac{\mathbf{v} \times \mathbf{h}}{\mu} - \hat{\mathbf{r}}$

5. $\omega = \text{ATAN2}\left(\dfrac{e_z \hat{n}_y}{\hat{h}_x},\ \hat{n}_x e_x + \hat{n}_y e_y\right)$

6. $\theta = \text{ATAN2}\left(\dfrac{r_y \hat{e}_x - r_x \hat{e}_y}{\hat{h}_x},\ r_x \hat{e}_x + r_y \hat{e}_y + r_z \hat{e}_z\right)$

The most notable differences from AL1 are steps 5 and 6 (calculation of $\omega$ and $\theta$). Because the angle between $\hat{\mathbf{n}}$ and $\mathbf{e}$ is $\omega$, and both vectors lie on the plane of the orbit, the properties of the dot and cross products imply:

$$\hat{\mathbf{n}} \cdot \mathbf{e} = e\cos(\omega), \tag{3.1}$$

$$\hat{\mathbf{n}} \times \mathbf{e} = e\sin(\omega)\hat{\mathbf{h}}, \tag{3.2}$$

where the fact that $\hat{\mathbf{h}}$ is the unit normal to the orbital plane has been used. Note that (3.1) and the magnitude of (3.2) can be thought of as the components of $\mathbf{e}$ in a base $\{\hat{\mathbf{n}}, \hat{\mathbf{b}}\}$ of the orbital plane such that $\{\hat{\mathbf{n}}, \hat{\mathbf{b}}, \hat{\mathbf{h}}\}$ is a right-handed triad. That is, $\hat{\mathbf{b}} = \hat{\mathbf{h}} \times \hat{\mathbf{n}}$ is obtained rotating $\hat{\mathbf{n}}$ $+90^\circ$ about $\hat{\mathbf{h}}$. Vector $\hat{\mathbf{b}}$ shall prove useful to derive the expressions of the branchless algorithm. Expanding the components of the cross product in (3.2):

$$\hat{\mathbf{n}} \times \mathbf{e} = \left(e_z \hat{n}_y,\ e_z \hat{n}_x,\ e_y \hat{n}_x - e_x \hat{n}_y\right). \tag{3.3}$$

Equating $x$ components of both sides of (3.2) yields

$$e\sin(\omega)\hat{h}_x = e_z \hat{n}_y. \tag{3.4}$$

Step 5 of the algorithm follows trivially from (3.1) and (3.4). Unfortunately, this procedure breaks down for equatorial orbits, because the vector $\hat{\mathbf{n}}$ is undefined. Likewise, to determine the true anomaly, start with:

$$\hat{\mathbf{e}} \cdot \mathbf{r} = r\cos(\theta), \tag{3.5}$$

$$\hat{\mathbf{e}} \times \mathbf{r} = r\sin(\theta)\hat{\mathbf{h}}, \tag{3.6}$$

and follow, *mutatis mutandi*, the calculation of $\omega$. In this case, the method fails if the orbit is circular, causing $\hat{\mathbf{e}}$ to become undefined.

More variations are possible manipulating the equations of Keplerian motion. For example, it is possible to avoid the explicit calculation of the eccentricity vector. Using the expression of the radial distance to the primary:

$$e\cos(\theta) = \frac{p}{r} - 1, \qquad (3.7)$$

where $p = h^2/\mu$ is the *semilatus rectum*. From the radial velocity:

$$e\sin(\theta) = \frac{\mathbf{v}\cdot\mathbf{r}}{\mu r h}. \qquad (3.8)$$

Combining equations (3.7) and (3.8) yields the eccentricity and the true anomaly:

$$e = \sqrt{\left(\frac{\mathbf{v}\cdot\mathbf{r}}{\mu r h}\right)^2 + \left(\frac{p}{r} - 1\right)^2}, \qquad (3.9)$$

$$\theta = \mathrm{ATAN2}\left(\frac{\mathbf{v}\cdot\mathbf{r}}{\mu r h}, \frac{p}{r} - 1\right). \qquad (3.10)$$

Next, $\omega$ can be determined via the argument of longitude $l = \omega + \theta$:

$$\hat{\mathbf{n}}\cdot\mathbf{r} = r\cos(l), \qquad (3.11)$$

$$\hat{\mathbf{n}}\times\mathbf{r} = r\sin(l)\hat{\mathbf{h}}, \qquad (3.12)$$

and so forth. Again, this method breaks down for circular orbits (both arguments of the arctangent in (3.10) are null, irrespective of the value of $\theta$) or if the inclination vanishes ($\hat{\mathbf{n}}$ in (3.11) and (3.12) is undefined).

The procedures outlined above share with AL1 the limitation of relying (implicitly or explicitly) on the directions $\hat{\mathbf{n}}$ and $\hat{\mathbf{e}}$ to compute some of the elements. Thus, they require a separate treatment when these unit vectors become undefined.

For the sake of brevity, further discussions shall focus on AL2 (AL1 is not really usable in practice, as discussed in the previous section) because it is the most popular approach in the literature. It is unfeasible to cover all the strategies in use for calculating the COEs, and their performance and limitations tend to be very similar. Furthermore, the changes to AL2 that yield an efficient and accurate branchless scheme can be ported easily to alternative algorithms, such as those discussed in this section.

### 3.2 The branchless algorithm

To understand the development of the branchless scheme, is vital to know that ATAN2, as specified by IEEE-754 (IEE, 2019), is not just a shortcut for computing $\arctan(y/x)$ and then adjusting the quadrant. Besides returning the correct value when $x = 0$, the result is well-defined and predictable even if both arguments are null ($0$ or $\pi$, depending on the signs of the operands[5]). ATAN2 implementations compliant with IEE-754 also guarantee high precision. For example, if the argument of $x + iy$ is close to $\pi/2$

---

[5] In standard floating-point arithmetic, zeros are signed.

, the computer may choose to evaluate $\pi/2 - \arctan(x/y)$ to preserve accuracy. ATAN2 can help compute $i$, $\Omega$ and the direction of the eccentricity vector to the best accuracy achievable by the hardware, free from exceptions and without having to set arbitrary tolerances. Moreover, in systems where it is a native instruction, it tends to execute faster than a software implementation of the same functionality. For systems without adequate ATAN2 support, Section 5 presents an alternative scheme that retains many benefits of the fully-branchless algorithm, but only uses the arccosine function.

The first step to remove the branches is finding a universal method of determining the direction of the ascending node, irrespective of the orbital inclination. This is trivial in retrospect, compute $\Omega$ first and then determine $\hat{\mathbf{n}}$:

$$\hat{\mathbf{k}} \times \mathbf{h} = -h_y \hat{\mathbf{i}} + h_x \hat{\mathbf{j}} \rightarrow \Omega = \mathrm{ATAN2}(h_x, -h_y) \rightarrow \hat{\mathbf{n}} = \cos\Omega \hat{\mathbf{i}} + \sin\Omega \hat{\mathbf{j}}. \qquad (3.13)$$

As discussed, ATAN2 gives the best approximation to $\Omega$ that the hardware is capable of. Even if both arguments vanish (i.e. the orbit is equatorial) the result remains valid, albeit arbitrary ($0$ or $\pi$). This is acceptable because, for an equatorial orbit, any point of the trajectory can be considered the ascending node. With a reliable value of $\Omega$, $\hat{\mathbf{n}}$ is uniquely determined in a consistent way. This simple change removes the need to treat inclined and equatorial orbits separately.

Next, to address circular orbits seamlessly, a robust reference frame of the orbital plane is required to avoid excessive reliance on $\hat{\mathbf{e}}$. This is just a matter of computing $\hat{\mathbf{b}} = \hat{\mathbf{h}} \times \hat{\mathbf{n}}$, which completes the triad $\{\hat{\mathbf{n}}, \hat{\mathbf{b}}, \hat{\mathbf{h}}\}$. The angle between the reference direction $\hat{\mathbf{n}}$ and a generic vector $\mathbf{a}$ contained in the orbital plane can be safely computed using $\mathrm{ATAN2}(\mathbf{a} \cdot \hat{\mathbf{b}}, \mathbf{a} \cdot \hat{\mathbf{n}})$. Thus:

$$\omega = \mathrm{ATAN2}(\mathbf{e} \cdot \hat{\mathbf{b}}, \mathbf{e} \cdot \hat{\mathbf{n}}), \qquad (3.14)$$

$$l = \mathrm{ATAN2}(\mathbf{r} \cdot \hat{\mathbf{b}}, \mathbf{r} \cdot \hat{\mathbf{n}}). \qquad (3.15)$$

In a circular orbit $\mathbf{e}$ vanishes, but (3.14) still returns a valid $\omega$ ($0$ or $\pi$). Either value is acceptable, because the pericenter is not defined. The calculation of the argument of longitude is always safe, because the position vector never vanishes in real scenarios (that would mean the orbiter has impacted the center of the primary). Finally, the true anomaly is simply $\theta = l - \omega$. The arbitrary choice of $\omega$ in circular orbits has no detrimental effect, because the sum $\theta + \omega$ always yields the correct angle between the reference direction and the orbiter.

The flowchart of the branchless algorithm (AL3 hereafter) is shown in Fig. 3. Clearly, the branchless algorithm is easier to follow and requires fewer steps. This simplicity is a considerable advantage by itself, as it reduces the likelihood of coding errors and improves maintainability of the software. AL3 also removes the need to set a threshold for eccentricity, further reducing the probability of implementation errors. In addition to eliminating performance-crippling branches, use of ATAN2 removes the need to check the magnitude of the arccosine argument, and determining the sign of the angle afterwards. Of course, it can be argued that some of these tests have not disappeared, they are simply hidden inside the ATAN2 call. However, when the function is supported by the CPU, these tasks are accomplished at the hardware level, which is usually more efficient than a software implementation.

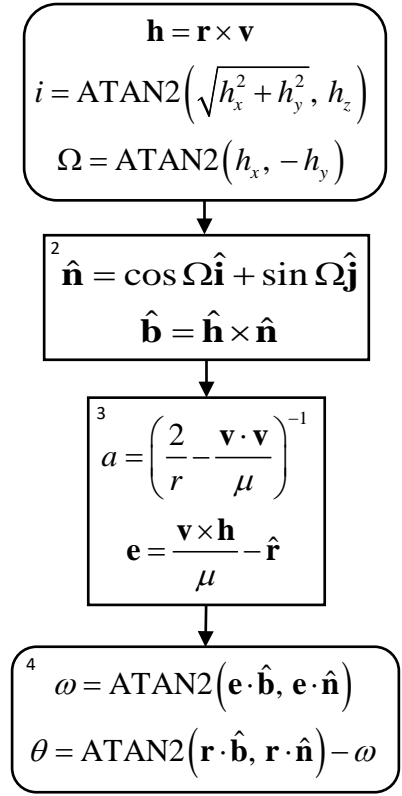

Fig. 3 - Branchless scheme (AL3).

Note that, in Step 1, the inclination is computed using the ATAN2 function instead of the arccosine. Replacing the arccosine with the arctangent improves the accuracy when operating in finite precision arithmetic. For small values of the inclination

$$\cos i \simeq 1 - \frac{i^2}{2}. \qquad (3.16)$$

In floating-point arithmetic, there exists $\varepsilon > 0$ such that[6]

$$\text{if } |x| < \varepsilon \Rightarrow 1 + x \triangleq 1, \qquad (3.17)$$

---

[6] Rigorously speaking, there are two alternative definitions of $\varepsilon$. The first one is given by Eq. (3.17). The second one is the relative rounding error of floating-point arithmetic operations, which is half as large. For the sake of simplicity, both of them shall be used interchangeably, because it does not affect the substance of the discussion.

where a hat over the equal sign indicates the result of approximate arithmetic (Golberg, 1991). For double precision, $\varepsilon \sim 10^{-16}$ (IEE, 2019). Combining (3.16) and (3.17),

$$\text{if } |i| \leq \sqrt{\varepsilon} \Rightarrow \cos i \simeq 1 - \frac{i^2}{2} \hateq 1 \Rightarrow \arccos(\hat{\mathbf{h}} \cdot \hat{\mathbf{k}}) \hateq 0. \qquad (3.18)$$

As the inclination of the orbit approaches $\sqrt{\varepsilon}$, AL2 yields vanishing inclination, introducing additional errors. Interestingly, while this phenomenon is undesirable from the accuracy standpoint, it has a convenient side effect. Because the computed value of the inclination vanishes well before the angular momentum becomes truly parallel to the $z$ direction, it is possible to apply directly the condition $i > 0$ (i.e., without setting a finite tolerance) to select the code path for equatorial/inclined orbits.

The ATAN2 instruction solves the accuracy issue because, for small inclinations,

$$\tan i \simeq i. \qquad (3.19)$$

There is no constant term summed to the inclination in Eq. (3.19). Therefore, the arctangent function computes a good numerical approximation of the inclination, no matter how small it is. This same advantage applies whenever the arccosine function is used to compute an angle close to 0 or $\pi$, so it also affects the determination of $\Omega$, $\omega$ and $\theta$. To keep the presentation concise, this paper focuses on the inclination, but the discussion also applies to the other angles.

It must be mentioned that AL3 is not truly general, as it will not work for zero orbital angular momentum. However, this case is of minimal relevance in practical applications (the trajectory is a straight line passing through the primary) and can usually be ignored.

## 4. Comparative accuracy of AL2 and AL3

The accuracy of the two implementations was tested building families of orbits with decreasing values of eccentricity, inclination, or both. Each orbit is sampled at $N = 100$ equally-spaced values of the true anomaly (starting at the pericenter). The error due to the transformation between state vector and orbital elements is estimated as follows:

1. From the reference orbital elements $\mathbf{o}_i^{ref}$ at the i-th sampling point, compute the corresponding reference state vector $\mathbf{x}_i^{ref} = \mathbf{x}(\mathbf{o}_i^{ref})$.
2. Apply the transformation to determine the approximate orbital elements $\tilde{\mathbf{o}}_i = \mathbf{o}(\mathbf{x}_i^{ref})$. Then, invert the transformation to recover an approximation of the reference state vector $\tilde{\mathbf{x}}_i^{ref} = \mathbf{x}(\tilde{\mathbf{o}}_i)$.
3. The absolute error at the i-th point is $\delta_i = \|\mathbf{x}_i^{ref} - \tilde{\mathbf{x}}_i^{ref}\|$.

For each orbit, the maximum error across all the sampling points and the RMS error are recorded:

$$\delta^{\max} = \max\{\delta_1,\ldots,\delta_N\} \quad ; \quad \delta^{RMS} = \sqrt{\frac{\sum_1^N \delta_i^2}{N}}. \tag{4.1}$$

This error measure includes a contribution from the transformation of orbital elements back to state vector. However, given that this part affects equally both AL2 and AL3, it yields valid comparisons. Note that it is not appropriate to evaluate the error by comparing $\mathbf{o}_i^{ref}$ directly against $\tilde{\mathbf{o}}_i$ because, near the singularities of the transformation, vastly different combinations of $\{\Omega, \omega, \theta\}$ correspond to almost identical state vectors. Instead, (4.1) measures how well the original state vector can be recovered from the approximate orbital elements. This is more relevant from a physical standpoint.

For the sake of simplicity, dimensionless variables have been used. In this system, the gravitational parameter is $\mu = 1$, and a circular orbit of unitary radius has an orbital velocity of one. The fixed values $a = 1$, $\Omega = \omega = 1$ rad have been used to focus on the effects of $e$ and $i$. Because the eccentricity is kept small in all tests (maximum 0.01) the distance to the primary and the velocity are always close to unity. Thus, the absolute and relative errors have the same order of magnitude and it is not necessary to examine them separately. The absolute error is used for the comparisons. All calculations use double precision floating-point arithmetic (i.e., $\varepsilon \sim 10^{-16}$).

## 4.1 Quasi-circular orbits with nonzero inclination

The inclination is fixed at $\pi/4$, and the eccentricity is progressively reduced from $10^{-2}$ to $10^{-16}$. The test compares the traditional scheme (AL2) against the branchless algorithm (AL3). For this initial run $e_{thr} = 0$, as in AL1. The results are shown in Fig. 4.

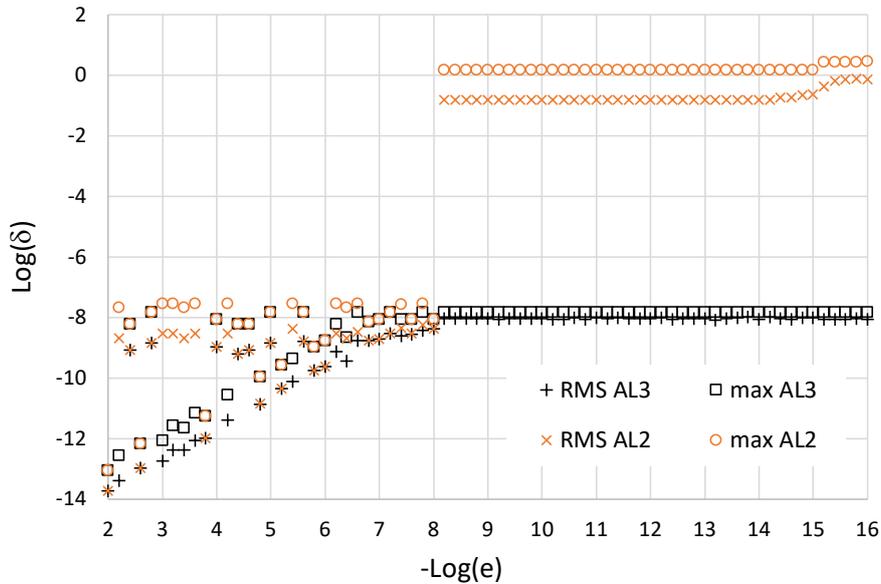

Fig. 4 – AL2 vs. AL3 error. Inclined quasi-circular orbits. $e_{thr}=0$.

As anticipated in section 2, the standard algorithm fails catastrophically for small eccentricities (below $10^{-8}$). The issue arises because, for quasi-circular orbits, the calculation of eccentricity involves the difference of two very similar vectors. It is a well-known fact in finite precision arithmetic that subtracting two close numbers is an ill-conditioned operation. The relative error of the result can be much larger than the uncertainties of the operands, a phenomenon known as catastrophic cancellation (Muller et al., 2018). It is easier to understand the source of the problem if, instead of the eccentricity vector, one focuses on its magnitude. The eccentricity can be recast as

$$e = \sqrt{1 - \frac{h^2}{\mu}\left(\frac{2\mu}{r} - v^2\right)}. \qquad (4.2)$$

With floating-point arithmetic, the evaluation of the radicand will be subject to a rounding error of order $O(\varepsilon)$ at least[7]. Therefore, the computed (approximate) eccentricity is

$$\tilde{e} = \sqrt{e^2 + O(\varepsilon)}. \qquad (4.3)$$

It is clear that, whenever $e < \sqrt{\varepsilon}$, the rounding error term dominates the calculation and $\tilde{e}$ becomes unreliable. This agrees perfectly with the behavior observed in Fig. 4, where AL2 becomes unusable for $e < 10^{-8} \approx \sqrt{\varepsilon}$. The reason for the unacceptable increase in $\delta$ is not the uncertainty of $e$ itself (the orbit is so close to circular that it does not change the result much), but the loss of accuracy of $\hat{\mathbf{e}}$ (the direction of the pericenter). This contaminates the calculation of $\omega$ and $\theta$, to the point that they correspond to a completely erroneous position along the orbit. AL3 is impervious to this issue because, by design, it ensures that the sum $\omega + \theta$ is the angle between the orbiter and the ascending node, even if $e \ll 1$.

AL2 can be fixed by treating the orbit as circular whenever the eccentricity is sufficiently small. From the discussion above, a reasonable threshold is $e_{thr} = 10\sqrt{\varepsilon} \approx 10^{-7}$. A safety factor 10 has been included in $e_{thr}$, to ensure that catastrophic failure never occurs.

---

[7] Note that the radicand is the difference of two operands, the first one having the fixed value 1. Therefore, for small eccentricities, the second approaches unity. The intrinsic rounding error of the floating-point approximation of a value close to 1 is $\varepsilon$. Therefore, the expected uncertainty of the radicand calculation is, at a minimum, $\varepsilon$.

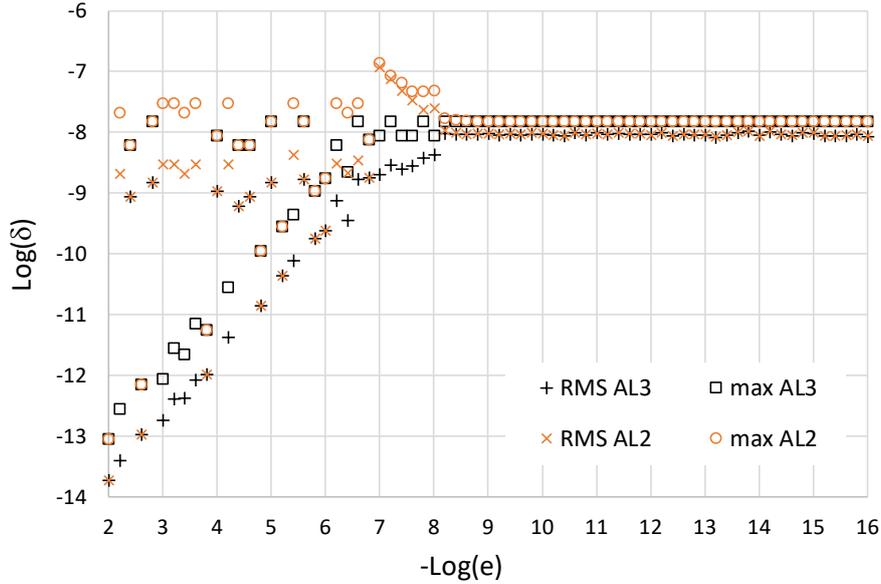

Fig. 5 – AL2 vs. AL3 error. Inclined quasi-circular orbits. $e_{thr}=10^{-7}$.

Adding the eccentricity tolerance makes the accuracy of AL2 and AL3 comparable (see Fig. 5). Note that, superimposed to the general trend, there are strong fluctuations of the error (by as much as 5 orders of magnitude). This is typical of rounding uncertainties, because they can cancel out or add up unpredictably. It is remarkable that, whenever there is a difference between both schemes, the error of AL3 is always lower. From this point on, all test involving AL2 use $e_{thr} = 10\sqrt{\varepsilon}$ .

### 4.2 Elliptical orbits with small inclination

In this case the eccentricity is fixed at 0.01 and the inclination varied from $10^{-2}$ to $10^{-16}$ rad.

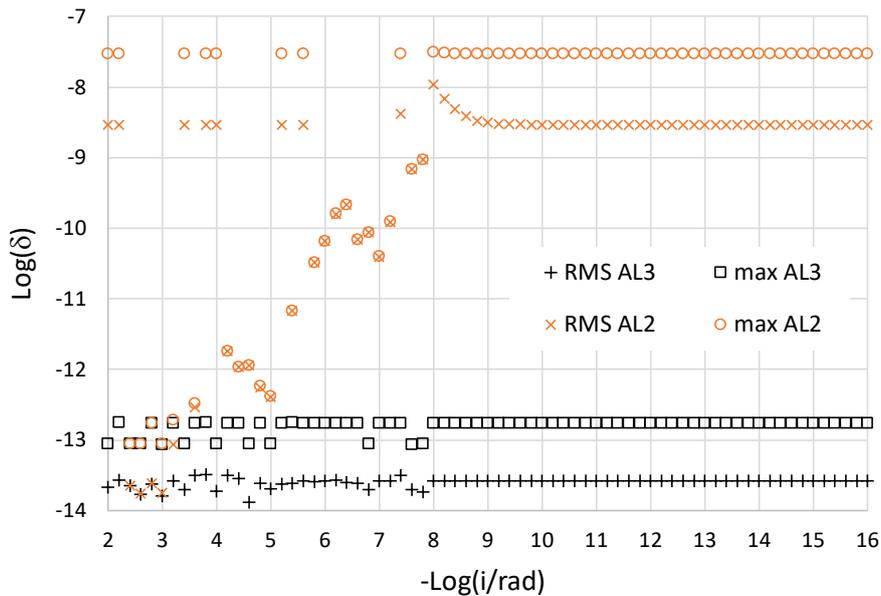

Fig. 6 - AL2 vs. AL3 error. Small-inclination elliptical orbits. $e_{thr}=10^{-7}$.

The vast superiority of the improved algorithm is evident in Fig. 6. The reason is the limitations of the arccosine function, illustrated in Eq. (3.18).

### 4.1 Orbits with vanishing eccentricity and inclination

Finally, eccentricity and inclination are set to the same value, which is ramped down from $10^{-2}$ to $10^{-16}$. The behavior of both schemes is very similar (see Fig. 7), because the uncertainty of the eccentricity dominates the error. Note that, mirroring the inclined low-eccentricity test, whenever there is a difference between the two schemes, AL3 delivers better accuracy.

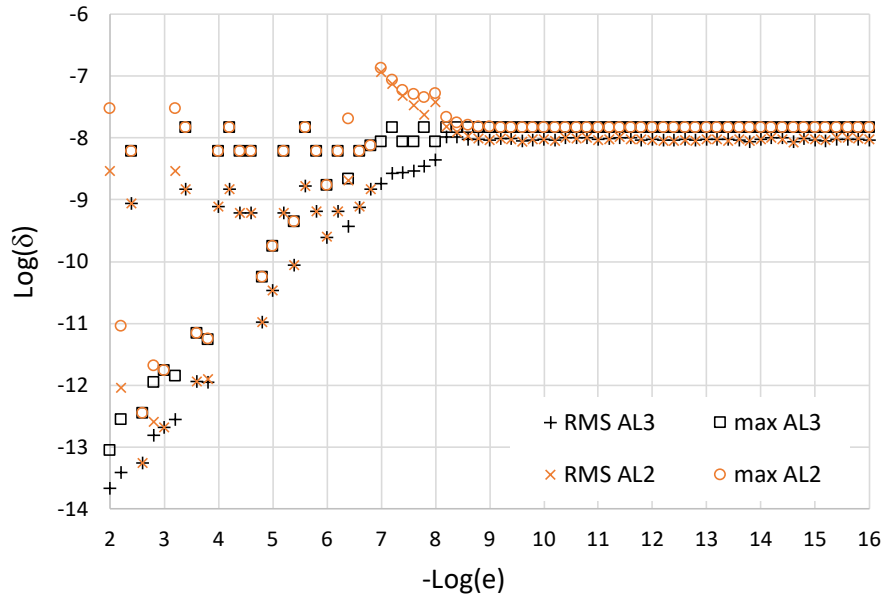

Fig. 7 - AL2 vs. AL3 error. Small eccentricity and inclination (e=i/rad). $e_{thr}=10^{-7}$.

### 5. A hybrid algorithm with reasonable performance and accuracy

The scheme AL3 is very accurate, robust and extremely simple to implement in comparison with AL2. However, the fact that it relies on calls to ATAN2 can be an issue in processors architectures with limited Floating-Point Unit (FPU) capabilities. This is often the case in microcontrollers and embedded systems. Those processors may not support the instruction natively.

In absence of hardware support for ATAN2, it must be emulated via software. For example, in the C language, including the *math.h* library makes ATAN2 accessible, irrespective of the CPU capabilities. Compilers that adhere strictly to the IEEE-754 standard will handle null arguments correctly, and AL3 will behave as expected (albeit with reduced performance due to the lack of dedicated hardware). However, proper IEE-754 support is not universal, and must be verified for each combination of hardware and compiler. In some systems, ATAN2 may trigger and exception when both arguments are zero. Even if ATAN2 is supported correctly, the performance of AL3 in systems with slow FPUs can be unsatisfactory. Moreover, for some applications, a universal solution that does not depend on the specifics of the hardware and software may be preferable.

To address those situations where AL3 is impractical, an alternative scheme that relies only on calls to the arccosine function (like AL2) is presented. Furthermore, it removes the two transcendental function evaluations (sine and cosine) that AL3 uses to compute $\hat{\mathbf{n}}$ (Step 2 in Fig. 3), which can be expensive in systems with weak FPUs[8].

There are two key insights for improving the performance of AL2 and retain some of the accuracy benefits of AL3, all the while avoiding calls to ATAN2. The first one is that the loss of accuracy of AL2 when the inclination decreases is due to the presence of the arccosine function (see section 3), which has zero slope at the origin. The second one is that, as shown in section 4.1, the calculated value of the eccentricity effectively turns into numerical noise well before the orbit becomes circular (i.e., for $e < \sqrt{\varepsilon}$ it becomes effectively irrelevant if the orbit is circular or not). Thus, it is not necessary to preserve accuracy as $e$ approaches zero. It is sufficient to guarantee that the result is reasonable.

It would be possible to remove ATAN2 from the computation of inclination and retain the full accuracy for small $i$ by using the arcsine. Thus:

$$i = \begin{cases} \arcsin\left(\dfrac{\sqrt{h_x^2 + h_y^2}}{h}\right) & \text{for small inclination} \\ \arccos\left(\dfrac{h_z}{h}\right) & \text{otherwise} \end{cases}. \quad (5.1)$$

Note that it is not appropriate to always use the arcsine, as its accuracy degrades for polar orbits[9]. This introduces one branch in the code, but brings several advantages to the table, as shown later. The condition "small inclination" in Eq. (5.1) must be defined precisely and in a robust way (i.e., not linked to any particular case). The optimal threshold for small inclinations is the point where the accuracies of the two branches of (5.1) become comparable, with a twist to increase performance.

The accuracy of the cosine branch can be estimated assuming the evaluation of the term $h_z/h$ is subject to an absolute error $\varepsilon$ (small inclination is assumed, so $h_z/h \approx 1$)

$$\frac{h_z}{h} \cong \cos i + \varepsilon . \quad (5.2)$$

Using the series expansion of the cosine, it is possible to estimate the computed value of the inclination ($\tilde{i}$)

$$\cos i + \varepsilon = \cos \tilde{i} = 1 - \frac{\tilde{i}^2}{2} + O(\tilde{i}^4). \quad (5.3)$$

---

[8] For processors with capable FPUs, this performance penalty is not as large as it seems. x86 processors support the FSINCOS instruction (Intel Corp., 2024). It computes simultaneously the sine and cosine of the argument, at a lower cost than two separate evaluations.

[9] This is another example of the advantages of ATAN2, where the processor itself chooses the most accurate way to evaluate the angle, depending on the values of the arguments.

Expanding the left-hand side (LHS)

$$-\frac{i^2}{2}+\varepsilon = -\frac{\tilde{i}^2}{2}+O(i^4), \quad (5.4)$$

where $i \approx \tilde{i}$ is assumed to lump all the higher-order terms together. Retaining only second-order terms gives an estimate of the absolute error of the inclination computed with the cosine ($\delta^{\cos}$):

$$2\varepsilon = |i^2 - \tilde{i}^2| = (i+\tilde{i})|i-\tilde{i}| \approx 2i\delta^{\cos} \rightarrow \delta^{\cos} \sim \frac{\varepsilon}{i}. \quad (5.5)$$

To estimate the error of the sine formula, consider the series expansion

$$\frac{\sqrt{h_x^2 + h_y^2}}{h} = \sin i = i - \frac{i^3}{6} + O(i^5). \quad (5.6)$$

To maximize performance, assume only the first term of the series is retained. In this case, computing inclination is trivial. The error of this simple approximation is

$$\delta^{\sin} \sim \frac{i^3}{6}. \quad (5.7)$$

The crossover point ($i_{cro}$) is the inclination which makes (5.5) and (5.7) comparable:

$$\frac{\varepsilon}{i_{cro}} \sim \frac{i_{cro}^3}{6} \rightarrow i_{cro} \sim \sqrt[4]{6\varepsilon}. \quad (5.8)$$

For double precision arithmetic, $i_{cro} \sim 10^{-4}$. This is ideal, because it matches the point where the cosine approximation accuracy starts to degrade rapidly (see Fig. 6). At the cost of one branch in the scheme, the ATAN2 call for the inclination has been replaced by an arccosine (for inclined orbits) or, even better, the identity function (for small *i*). Moreover, it is possible to recast the calculation of $\hat{\mathbf{n}}$ in a way that avoids additional calls to trigonometric functions:

$$\hat{\mathbf{n}} = \begin{cases} \dfrac{-h_y}{h_{xy}}\hat{\mathbf{i}} + \dfrac{h_x}{h_{xy}}\hat{\mathbf{j}} & \text{if } i > i_{thr} \\[2mm] \dfrac{-\overline{h}_y}{\overline{h}_{xy}}\hat{\mathbf{i}} + \dfrac{h_x}{\overline{h}_{xy}}\hat{\mathbf{j}} & \text{otherwise} \end{cases}, \quad (5.9)$$

where $h_{xy} = \sqrt{h_x^2 + h_y^2}$ and the overline denotes "safe" values. These are computed as

$$\overline{h}_y = \text{sign}\left(\max\left(|h_y|, \varepsilon h\right), h_y\right) \; ; \; \overline{h}_{xy} = \sqrt{\overline{h}_y^2 + h_x^2}. \quad (5.10)$$

The effect of the safe values is to make $\hat{\mathbf{n}}$ approach $\hat{\mathbf{i}}$ when $i \ll \varepsilon$ (it acts as a safeguard against division by zero). Note that

$$h_{xy} = h \sin i. \quad (5.11)$$

Therefore, the safe values (5.10) are extremely close to the original (non-overlined) variables, unless the inclination is very small. They have a negligible impact on the accuracy because, by the time the correction becomes important, the orbit is quasi-equatorial and the errors are dominated by the *xy* components of position and velocity. Under these conditions, the exact location of the ascending node is not crucial. Those readers concerned with extreme accuracy can replace $\varepsilon$ in Eq. (5.10) with a smaller value, but the effect is very limited in practice.

Moving to the efficient determination of $\{\Omega, \omega, \theta\}$, having already established a branch for inclined/equatorial orbits makes it possible to completely eliminate the calls to ATAN2.

Start by computing:

$$\Omega = \text{sign}\left(\arccos(\hat{n}_x), \hat{n}_y\right), \quad (5.12)$$

$$\omega_{aux} = \arccos\left(\frac{\mathbf{e} \cdot \hat{\mathbf{n}}}{e + \varepsilon}\right), \quad (5.13)$$

$$\xi = \max\left(\min\left(\hat{\mathbf{r}} \cdot \hat{\mathbf{n}}, 1\right), -1\right) \rightarrow \theta_{aux} = \arccos(\xi). \quad (5.14)$$

The $\varepsilon$ term in the denominator of Eq. (5.13) ensures that no exception occurs for very low eccentricity orbits ($\omega \rightarrow 0$ for quasi-circular orbits). Furthermore, it removes the need to check that the argument of the arccosine is below one. In practice, the extra term entails no loss of accuracy because, as discussed in section 4.1, the computation of eccentricity is subject to larger uncertainties due to its expression being numerically ill-conditioned.

For inclined orbits $(i > i_{thr})$, the signs of $\omega$ and $\theta$ are obtained from the vertical components of the eccentricity and position vectors:

$$\omega = \text{sign}(\omega_{aux}, e_z), \quad (5.15)$$

$$\theta = \text{sign}(\theta_{aux}, r_z) - \omega. \quad (5.16)$$

For small inclinations $(i < i_{thr})$, an auxiliary vector $\mathbf{b}$ determines the signs:

$$\mathbf{b} = h_z\left(\hat{\mathbf{k}} \times \hat{\mathbf{n}}\right) = h_z\left(-\hat{n}_y \hat{\mathbf{i}} + \hat{n}_x \hat{\mathbf{j}}\right), \quad (5.17)$$

$$\omega = \text{sign}(\omega_{aux}, \mathbf{e} \cdot \mathbf{b}), \quad (5.18)$$

$$\theta = \text{sign}(\theta_{aux}, \mathbf{r} \cdot \mathbf{b}) - \omega. \quad (5.19)$$

The vector $\mathbf{b}$ in Eq. (5.17) is contained in the *xy* plane. This simplifies the evaluation of the expressions where it appears (it only has two nonzero components), enhancing performance. It is orthogonal to $\hat{\mathbf{n}}$, with its direction determined by the sign of $h_z$. This ensures that the algorithm works also for retrograde orbits.

Note that, for the sake of clarity, the expressions above used $i > i_{thr}$ to identify inclined orbits, but the correct implementation is

$$\left|\frac{h_z}{h}\right| < \cos i_{cro}, \tag{5.20}$$

which addresses both prograde and retrograde trajectories.

The enhanced algorithm (AL4 henceforth) is summarized in Fig. 8. Note that the alternate code path for near-equatorial orbits (Steps 4-9) saves one arccosine evaluation, which is advantageous for the kind of systems targeted (i.e., weak FPUs).

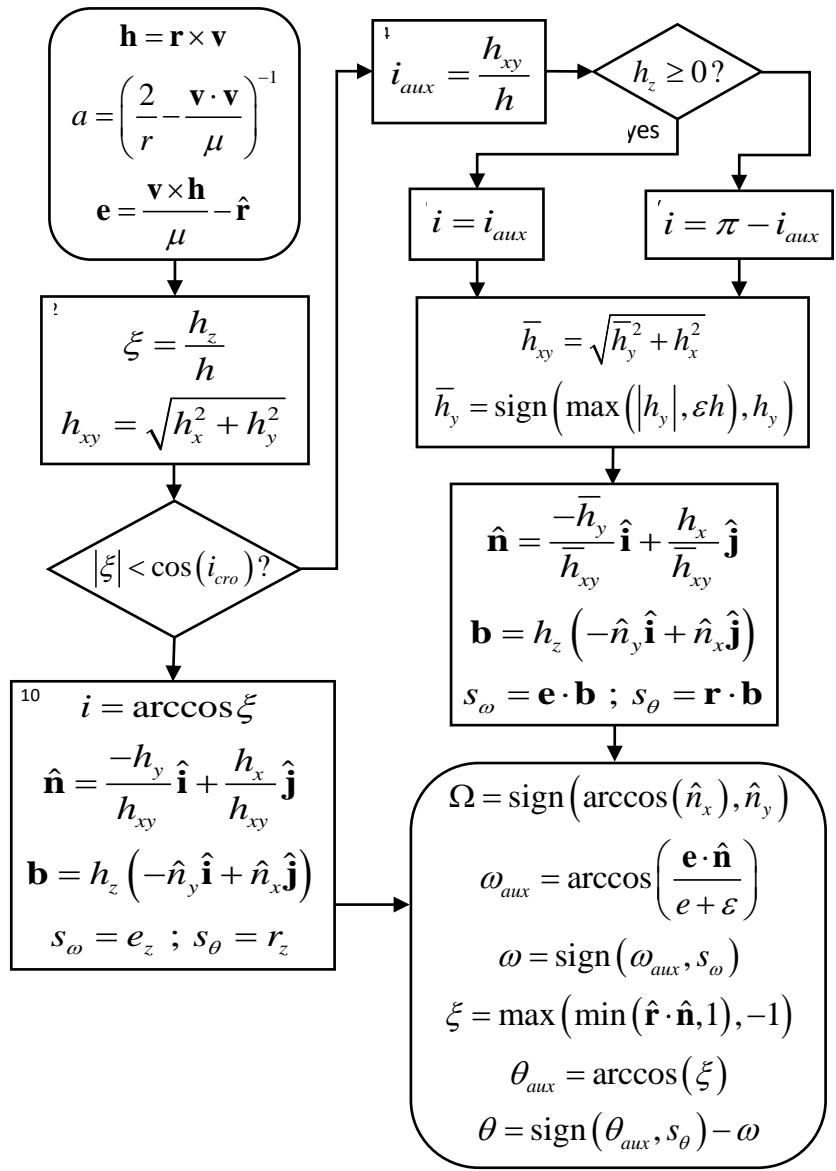

Fig. 8 - Hybrid efficient scheme (AL4).

## 5.1 Accuracy of AL4 vs. AL3

The only noticeable difference between AL3 and AL4 is in the elliptical small inclination test (Fig. 4). Because the other two benchmarks yield indistinguishable results, they will not be presented to save space.

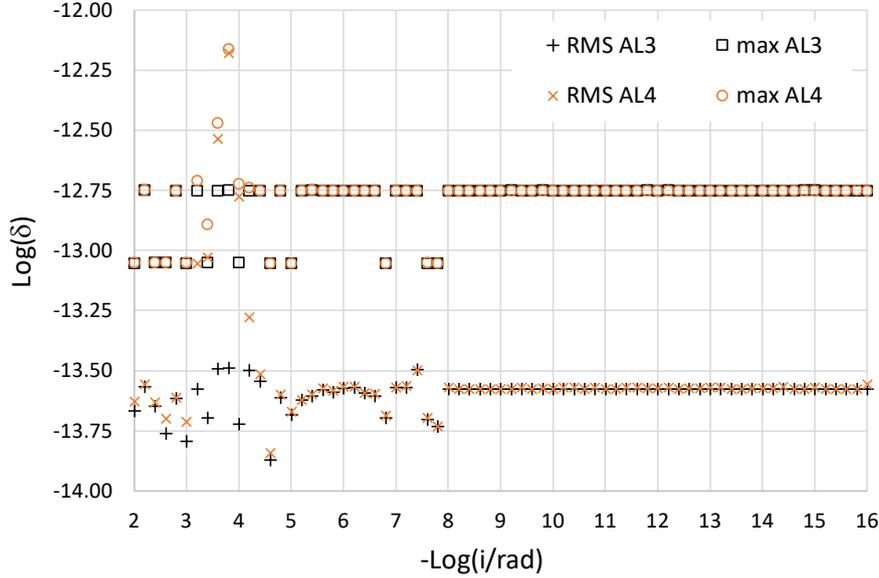

Fig. 9 – AL3 vs. AL4 error. Small-inclination elliptical orbits.

There is a moderate degradation of accuracy (up to one order of magnitude) in the neighborhood of $i_{cro}$, when the cosine approximation starts to lose precision and the degree-3 term of the sine series expansion is still relevant. The difference is subtle and can be ignored in most applications. If maximum accuracy is sought, AL4 can be improved by retaining one more term in the expansion of the sine. Starting from

$$\sin i = i - \frac{i^3}{3!} + \frac{i^5}{5!} + \cdots, \quad (5.21)$$

and dropping terms of order 5 and higher:

$$\frac{h_{xy}}{h} = i - \frac{i^3}{6}. \quad (5.22)$$

While there is an analytical solution for Eq. (5.22), its expression is computationally costly. It is more efficient to use a single Newton-Rapson iteration to compute the approximate inclination.

Let

$$f(i) = i - \frac{i^3}{6} - \frac{h_{xy}}{h} \rightarrow f'(i) = 1 - \frac{i^2}{2}. \quad (5.23)$$

The first iteration of the scheme is

$$i_0 = \frac{h_{xy}}{h}$$

$$i_1 = i_0 - \frac{f(i_0)}{f'(i_0)} = i_0 - \frac{-i_0^3}{6\left(1 - \frac{i_0^2}{2}\right)} = i_0 + \frac{i_0^3}{6 - 3i_0^2}. \qquad (5.24)$$

The resulting algorithm (denoted AL5 hereafter) requires only a minor change in Step 4 (see Fig. 10).

$$\xi = \frac{h_{xy}}{h}$$

$$i_{aux} = \xi + \frac{\xi^3}{6 - 3\xi^2}$$

Fig. 10 - Improved calculation of i for small-inclination orbits (AL5).

The crossover inclination for AL5 is determined using as error estimate for the sine approximation the degree-5 term of the series (5.21)

$$\frac{\varepsilon}{i_{cro}} \sim \frac{i_{cro}^5}{5!} \rightarrow i_{cro} \sim \sqrt[6]{120\varepsilon}, \qquad (5.25)$$

which yields $i_{cro} \sim 5 \cdot 10^{-3}$ for double precision. The accuracy of AL5 in the inclination test is virtually identical to AL3, so the graph is not included for the sake of brevity.

## 6. Additional benchmarks

This section extends the comparison to orbits with a wide range of orbital elements chosen at random. The algorithms AL2 through AL5 (AL1 is unusable in practice, so it is not included) were tested on two sets of one billion ($10^9$) combinations. The first set (termed "general" hereafter) uses uniformly distributed values within the following intervals:

$$a \in \left[10^{-3}, 10^3\right], \; e \in [0, 0.9], \; i \in [0, \pi], \; \{\Omega, \omega, \theta\} \in [0, 2\pi]. \qquad (6.1)$$

For the second set ("low e/i"), focusing on quasi-circular low-inclination orbits, the ranges of the orbital elements are:

$$a \in \left[10^{-3}, 10^3\right], \; \{\log e, \log i\} \in [-16, -2], \; \{\Omega, \omega, \theta\} \in [0, 2\pi]. \qquad (6.2)$$

Note that in (6.2) it is the logarithms of the eccentricity and inclination that are uniformly distributed. This ensures that all the orders of magnitude are represented equally in the sample. The gravitational parameter is 1 in all cases.

### 6.1 Comparative accuracy tests

This benchmark uses the relative error because the components of the state vector are no longer of order 1. For each individual in the sample, it is computed as

$$\varphi_i = \frac{\delta_i}{\left\|\mathbf{x}_i^{ref}\right\|}, \tag{6.3}$$

where the absolute error $\delta_i$ is given by Eq. (4.1). Table 1 summarizes the results.

| Orbit type | $\varphi$ | AL2 | AL3 | AL4 | AL5 |
|---|---|---|---|---|---|
| General | RMS | 3.55·10⁻¹¹ | 7.14·10⁻¹⁴ | 2.13·10⁻¹² | 2.13·10⁻¹² |
| | max | 2.76·10⁻⁰⁷ | 1.69·10⁻⁰⁹ | 2.73·10⁻⁰⁸ | 2.73·10⁻⁰⁸ |
| Low e/i | RMS | 1.90·10⁻⁰⁸ | 1.80·10⁻¹⁰ | 1.82·10⁻¹⁰ | 1.82·10⁻¹⁰ |
| | max | 2.98·10⁻⁰⁷ | 2.10·10⁻⁰⁸ | 2.28·10⁻⁰⁸ | 2.28·10⁻⁰⁸ |

*Table 1 – Relative accuracy comparison for random orbital elements. Sample size $10^9$.*

For the general set, AL3 excels at accuracy. It performs over one order of magnitude better than AL4 and AL5, whose behaviors are virtually identical. AL2 yields the largest error, more than two orders of magnitude worse than AL3. In the context of the low e/i set, AL3 through AL5 perform very close to each other, with an average improvement of two orders of magnitude over AL2. In this case, however, the peak difference is smaller, at one order of magnitude approximately. The tests in section 4.2 show that the difference can be much larger for some specific conditions (see Fig. 6, where the accuracy of AL2 is up to 5 orders of magnitude worse). However, those are unlikely to appear in a random set.

## 6.2 Comparative performance tests

Finally, the speed of the four algorithms in both the general and low e/i configurations is assessed. For the test, the sample sizes are reduced to $10^5$ combinations of orbital elements, and the corresponding state vectors stored in memory. Then, the transformation to orbital elements is applied to each set 1000 times (i.e., a total of $10^8$ transformations for each algorithm and set). To obtain a reliable timing, the entire process is repeated 10 times, averaging the duration of each pass. The set of $10^5$ states gives reasonable diversity while fitting easily inside the CPU cache. This improves the estimate of raw algorithm performance, by avoiding memory access bottlenecks.

The measurements were taken on an AMD Ryzen 9 7945HX laptop CPU using the Intel IFX Fortran compiler (version 2024.1) for Windows 64bit with the maximum optimization level (O3). The benchmark results are summarized in Table 2. The run times for each combination of orbit type and algorithm are normalized with respect to AL2 in the general case.

| Orbit type | AL2 | AL3 | AL4 | AL5 |
|---|---|---|---|---|
| General | 1.00 | 0.57 | 0.70 | 0.70 |
| Low e/i | 0.76 | 0.74 | 0.74 | 0.72 |

*Table 2 - Relative time comparison for general and quasi-circular low-inclination orbits.*

Code performance is affected by many implementation details. Changing the hardware, operating system, compiler version or settings can have a noticeable impact on speed. Furthermore, the effect may be different for each algorithm. The results from Table 2 should be taken as rough performance guidelines. It is always important to repeat this test on the target system, and select the scheme that best serves the requirements of the application at hand.

Keeping in mind the caveat of the previous paragraph, the performance of AL3 for general orbits is outstanding, requiring 43% less time than AL2. This exemplifies the large performance gains obtained by removing branches from the code. AL4 and AL5 deliver a 30% reduction compared with the traditional scheme.

For the low e/i set, all schemes achieve comparable speed. Interestingly, the performance of AL3 degrades substantially relative to the general case. This is probably due to the large variation in latency of the ATAN2 function between the best and worst-case scenarios. As an example, for the processor used in the test (AMD Zen 4 family), the latency of FPATAN varies from 50 to 190 cycles (Fog, 2024). It is likely that, when the protections against small values of the arguments are triggered, the latency of each ATAN2 call increases substantially. In any case, this is not an issue because, even in this situation, the speed is identical to AL4. While AL5 seems to fare slightly better[10], the difference with respect to AL3/4 is close to the uncertainty of the time measurements (around 1%), so the advantage is marginal at best. Another noteworthy fact is that AL2 behaves much better than in the general case. This is due to the alternative paths for low eccentricity and inclination being computationally simpler (they just assign arbitrary fixed values to the ill-defined variables). This offsets the penalties due to the multiple branches, and brings the performance close to the other algorithms.

6.3 Comparison summary

The benchmarks in sections 4 and 6 evidence that, in systems with an efficient implementation of ATAN2, AL3 is the undisputed choice. It delivers the best performance, accuracy and ease of implementation. If AL3 is not feasible, AL5 is the obvious alternative. It is as fast as AL4 (even displaying a slight advantage in the low e/i test), equally simple to implement, and can improve accuracy for some low inclination orbits, as indicated in section 5.1. However, it is important to keep in mind that the differences between AL5 and AL4 are very limited in practice. Regarding AL2, it delivers the worst performance in all areas.

7. Conclusions

This paper reviewed the standard approach to computing classical orbital elements from spacecraft state vector. Due to the singularities that affect the orbital elements in the case of circular and zero-inclination orbits, they must be addressed separately. The algorithm, if implemented in the way commonly presented in the literature (scheme AL1), suffers from accuracy and efficiency shortcomings. The accuracy can be improved to a certain extent by increasing the complexity of the implementation (AL2), at the cost of performance. An important contributor to the computational cost of AL2 is the presence of branches in the code, which is detrimental for the pipelined architectures of modern processors.

---

[10] The small speed advantage relative to AL4 in this test is likely due to $i_{cro}$ being larger for AL5. This allows AL5 to take the code branch for small inclinations more frequently, avoiding an arccosine call.

The intrinsic safeguards of the ATAN2 instruction, available in most programming languages, enable a coding scheme free from branches (AL3). This approach also improves the accuracy of the transformation compared to AL2, reducing the errors by as much as 5 orders of magnitude for some small inclination orbits. To address systems with limited support for transcendental functions, a hybrid of the AL2 and AL3 schemes has been developed (AL5) that avoids ATAN2 calls while retaining an important part of the performance and accuracy gains of AL3.

A test on a large set of random orbits showed that AL3 is, on average, 2 orders of magnitude more accurate than AL2. The advantage of AL5 is smaller, but still substantial at one order of magnitude. Regarding computational performance, the cost of AL3 is up to 43% less than AL2 for a random mix of all types of orbits. The advantage of AL5 is lower, at 30%. When restricted to low-inclination quasi-circular orbits, the speed gains of AL3 and AL5 are very limited, but they retain a notable accuracy advantage.

The scheme AL3 yields important benefits in terms of simplicity (improving ease of programming and maintainability), accuracy and speed, with no obvious downsides. It is therefore an excellent alternative to the standard implementation (AL2). For systems with limited support for ATAN2, the scheme AL5 offers the same advantages, albeit to a lesser degree.

ACKNOWLEDGMENTS

The authors acknowledge Khalifa University of Science and Technology's internal grant CIRA-2021-65/8474000413 and project ELLIPSE/8434000533 funded by Abu Dhabi's Technology Innovation Institute. EF has been partially supported by the Spanish Ministry of Science and Innovation under projects PID2020-112576GB-C21 and PID2021-123968NB-I00.

REFERENCES

Arsenault, J. L., Ford, K. C. & Koskela, P. E., 1970. Orbit Determination Using Analytic Partial Derivatives of Perturbed Motion. AIAA Journal, Volume 8, pp. 4-12.
DOI 10.2514/3.5597

Bate, R. R., Mueller, D. D. & White, J. E., 1971. Fundamentals of Astrodynamics, pp. 61-64. New York: Dover Publications.
ISBN 978-0-486-60061-1

Battin, R. H., 1999. An Introduction to the Mathematics and Methods of Astrodynamics, Chap. 10. Reston, VA: AIAA.
ISBN 1-56347-342-9

Broucke, R. & Cefola, P., 1972. On the equinoctial orbit elements. Celestial Mechanics, Volume 5, pp. 303–310.
ISBN 978-0-486-64687-9

Chobotov, V. A., 2002. Orbital Mechanics (3rd ed.), Chaps. 3 & 14. Reston, VA: AIAA.
ISBN: 978-1-56347-537-5

Cohen, C. J. & Hubbard, E. C., 1962. A Nonsingular Set of Orbital Elements. Astronomical Journal, 67(1), pp. 10-15.
DOI 10.1086/108597

Curtis, H. D., 2020. Orbital Mechanics for Engineering Students (4th ed.), pp. 191-193. Elsevier.
ISBN 978-0-08-102133-0

Danby, J. M. A., 1992. Fundamentals of Celestial Mechanics (2nd ed.), pp. 204-206. Richmond, VA: William-Bell.
ISBN 0-943396-20-4

Eyerman, S., Smith, J. E. & Eeckhout, L., 2006. Characterizing the branch misprediction penalty, pp. 48-58. 2006 IEEE International Symposium on Performance Analysis of Systems and Software, Austin, TX.

Fantino, E., Burhani, B., Flores, R. et al., 2023. End-to-end trajectory concept for close exploration of Saturn's Inner Large Moons. Communications in Nonlinear Science and Numerical Simulation, Volume 126, 107458.
DOI 10.1016/j.cnsns.2023.107458

Fantino, E., Flores, R. M., di Carlo, M. et al., 2017. Geosynchronous inclined orbits for high-latitude communication. Acta Astronautica, Issue 140, pp. 570-582.
DOI 10.1016/j.actaastro.2017.09.014

Fog, A., 2024. Software optimization resources: Instruction tables. [Online]
Available at: https://www.agner.org/optimize/instruction_tables.pdf
[Accessed 13 Aug. 2024]

Giacaglia, G. E. O., 1977. The Equations of Motion of an Artificial Satellite in Nonsingular Variables. Celestial Mechanics, 15(2), pp. 191-215.
DOI 10.1007/BF01228462

Golberg, D., 1991. What Every Computer Scientist Should Know About Floating-Point Arithmetic. ACM Computing Surveys, 23(1), pp. 5-49.
DOI 10.1145/103162.103163

Hintz, G. R., 2008. Survey of Orbit Element Sets. Journal of Guidance, Control, and Dynamics, 31(3), pp. 785-790.
DOI 10.2514/1.32237

IEEE (Institute of Electrical and Electronics Engineers), 2019. IEEE Std 754-2019: IEEE Standard for Floating-Point Arithmetic.26          Manuscript submitted to Adv. Sp. Res.

Intel Corp., 2024. Intel® 64 and IA-32 Architectures Software Developer's Manual, vol. 2: Instruction Set Reference
Available: https://www.intel.com/content/www/us/en/developer/articles/technical/intel-sdm.html
[Accessed 13 Aug. 2024]

Lagrange, J. L., 1781. Théorie des variations séculaires des éléments des planètes. Première partie contenant les principes et les formules générales pour déterminer ces variations. Berlin: Nouveaux mémoires de l'Académie royale des sciences et belles-lettres de Berlin.

Lara, M., San-Juan, J. F. & López-Ochoa, L. M., 2014. Efficient semi-analytic integration of GNSS orbits under tesseral effects. Acta Astronautica, Volume 102, pp. 355-366.
DOI 10.1016/j.actaastro.2013.11.006

Moulton, F. R., 1970. An Introduction to Celestial Mechanics (2nd revised ed.). New York: Dover Publications.
ISBN 978-0-486-64687-9

Muller, J. M., Brunie, N., Dinechin, F. D. et al., 2018. Handbook of Floating-Point Arithmetic, p. 102. Cham, Switzerland: Birkhüser.
ISBN 978-0-8176-4705-6

Nacozy, P. E. & Dallas, S. S., 1977. The Geopotential in Nonsingular Orbital Elements. Celestial Mechanics, 15(4), pp. 453-466.
DOI 10.1007/BF01228611

Pikus, F. G., 2021. The Art of Writing Efficient Programs, pp. 103-110. Birmingham, UK: Packt Publishing Limited.
ISBN 9781800208117

Proietti, S., Flores, R., Fantino, E. et al., 2021. Long-term orbit dynamics of decommissioned geostationary satellites. Acta Astronautica, Volume 182, pp. 559-573.
DOI 10.1016/j.actaastro.2020.12.017

Prussing, J. E. & Conway, B. A., 2013. Orbital Mechanics, pp. 50-52. New York: Oxford University Press.
ISBN 978-0-19-983770-0

Shen, J. P. & Lipasti, M. H., 2005. Modern processor design: fundamentals of superscalar processors, pp. 39-54. Boston: McGraw-Hill Higher Education.
ISBN 978-0071230070

Smith, J., 1981. A study of branch prediction strategies. ISCA 81: Proceedings of the 8th annual symposium on Computer Architecture, pp. 135-148. Minneapolis, MN.

Tewari, A., 2007. Atmospheric and Space Flight Dynamics, pp. 120-121. Boston, MA: Birkhäuser.
ISBN 978-0-8176-4437-6

Vallado, D. A. & McClain, W. D., 2007. Fundamentals of Astrodynamics and Applications, pp. 119-122. Hawthorne, CA: Microcosm Press.
ISBN 978-1-881883-14-2

Walker, M., Ireland, B. & Owens, J., 1985. A set modified equinoctial orbit elements (Erratum). Celestial Mechanics, Volume 36, pp. 409-419.
DOI 10.1007/BF01238929

Walker, M. J. H., 1986. A Set of Modified Equinoctial Orbit Elements. Celestial Mechanics and Dynamical Astronomy, 38(4), pp. 391-392.
DOI 10.1007/BF01227493